\documentstyle[12pt,epsfig]{article}

\begin{document}
\begin {center}
{\bf {\Large
Pion exchange interaction in the $\gamma p \to p e^+e^-$ reaction} }
\end {center}
\begin {center}
Swapan Das \footnote {email: swapand@barc.gov.in} \\
{\it Nuclear Physics Division,
Bhabha Atomic Research Centre,  \\
Trombay, Mumbai-400085, India  }  \\
{\it Homi Bhabha National Institute, Anushakti Nagar, 
Mumbai-400094, India  }
\end {center}

\begin {abstract}
The $\rho^0-\omega$ interference has been studied in the dilepton 
invariant mass distribution spectra in the photonuclear reaction, but that
is not done for the gamma-nucleon reaction.
Recent
past, the $e^+e^-$ invariant mass distribution spectrum
in the $\gamma p$ reaction, i.e., $\gamma p \to pe^+e^-$ reaction, was
measured at Jefferson Laboratory to look for the $\rho^0 -\omega$
interference in the multi-GeV region.
To
study the mechanism of this reaction, the differential cross section of the
$e^+e^-$ invariant mass distribution is calculated in the quoted energy
region.
The
reaction is assumed to proceed as $\gamma p \to Vp$; $V \to e^+e^-$, where
$V$ denotes a vector meson, i.e., either $\rho^0$ or $\omega$ meson.
The
photoproduction of the vector meson is described by the Vector Meson
Dominance (VMD) model which consists of diagonal and off-diagonal processes.
The
diagonal process is described as $\gamma \to V; ~Vp \to Vp$. The low energy
$\omega$ meson photoproduction data is well described by the off-diagonal
process
which
is illustrated as $\gamma \to \rho^0; ~\rho^0 p \to \omega p$. The
reaction $\rho^0p \to \omega p$ proceeds due to one pion exchange 
interaction.
The
differential cross sections of the $\gamma p \to pe^+e^-$ reaction due to
the above processes of VMD model are compared, and the significance of the
pion exchange interaction is investigated in the energy region of
$\gamma$ beam available at Jefferson Laboratory.
\end {abstract}


PACS number(s): 13.60.-r, 13.60.Le, 13.20.-v

\section{Introduction}

The quantum interference of the $\rho^0$ and $\omega$ mesons occurs because
of the small difference between the masses of these mesons. This
interference is seen in the decay channels of the above mesons which are
produced in the nuclear and particle reactions in the multi-GeV region
\cite{bauer}.
The
$\rho^0$ meson is preferably identified by the dipion production since it
dominantly ($\sim 100\%$) decays as $\rho^0 \to \pi^+\pi^-$ \cite{behr0}.
Though the branching ratio of $\omega \to \pi^+\pi^-$ is very small
compared to that of $\rho^0$ meson, the $\rho-\omega$ interference is 
visible in the $\pi^+\pi^-$ production in the $e^+e^-$ \cite{conn} (see the
references there in) and $\gamma$ induced \cite{biggs2} reactions.
The
quoted interference is shown to occur in the semihadronic decay channel,
i.e., $\rho^0, \omega \to \pi^0\gamma$, in the photonuclear reaction
\cite{das1}.

The distinct $\rho -\omega$ interference has been reported in the dilepton
$e^+e^-$ production in the photonuclear reaction by Alvensleben et al.,
\cite{alven} and Biggs et al., \cite{biggs}. That was not studied in the
photonucleon reaction, only the information about it exits near the
threshold region \cite{lutz}.
Recent
past, the $e^+e^-$ invariant mass distribution spectrum in the $\gamma p$
reaction (i.e., $\gamma p \to pe^+e^-$ reaction) was measured at Jefferson
Laboratory (JLab) to the search the $\rho -\omega$ interference using
Bremsstrahlung photon beam of energy range 1.2 to 5.4 GeV. The preliminary
results are reported in Ref.~\cite{djal}.

The $\gamma p \to pe^+e^-$ reaction is assumed to proceed as
$\gamma p \to Vp$; $V \to e^+e^-$, where $V$ denotes the vector meson, i.e.,
either $\rho^0$ or $\omega$ meson.
The
photoproduction of these mesons can be described by the vector meson
dominance (VMD) model \cite{bauer}. According to that, a physical photon
is composed of bare photon $\gamma_B$ and its hadronic component
$\gamma_V$, i.e.,
\begin{equation}
|\gamma> \simeq
\left [ 1 -
 \sum_{V = \rho^0, \omega, \phi, ...}
 \frac{ \pi \alpha_{em} }{ 2\gamma^2_{\gamma V} } \right ] |\gamma_B>
+\sum_{V = \rho^0, \omega, \phi, ...}
 \frac{\sqrt{\pi \alpha_{em}}}{\gamma_{\gamma V}} |\gamma_V>,
\label{wgma}
\end{equation}
where $\alpha_{em} ( =1/137.04)$ is the fine structure constant.
$\gamma_{\gamma V}$ is the photon to vector meson coupling constant
\cite{saku}. The values of $\gamma_{\gamma \rho}$ and
$\gamma_{\gamma \omega}$ (as extracted from the measured width of
$\rho^0,\omega \to e^+e^-$ \cite{olive}) are 2.48 and 8.53 respectively
\cite{sibir1}. Since the cross section is calculated in the pole mass region
of the $\rho$ and $\omega$ mesons, other vector mesons (e.g., $\phi$ meson)
are not considered in the present context.

The hadron production in the nuclear and particle reactions occurs in the
GeV region. Since the strength of the electromagnetic interaction is much
less $(\sim 10^{-2})$ than that of the hadronic interaction, the first term
in Eq.~(\ref{wgma}) can be neglected in the quoted energy region.
The
latter, i.e., $\gamma_V$, appears as vector meson $V$ (e.g., $\rho^0$,
$\omega$, ... etc.) due to the photon nucleon (or nucleus) scattering in
the GeV region. This mechanism is illustrated by the diagonal and
off-diagonal processes of the vector meson dominance (VMD) model.
According
to the diagonal process, the hadronic part of the photon, i.e., $\gamma_V$
in Eq.~(\ref{wgma}), converts to virtual $V$ meson followed by the elastic
scattering,
i.e.,
$\gamma \to V$; $Vp \to Vp$. The coherent contribution of $\rho^0$ and
$\omega$ mesons, and the $\rho^0 -\omega$ interference can be described by
this process.
The off-diagonal process should not be considered for this purpose, since
the $\rho$ meson photoproduction in the off-diagonal process can be ignored
as it is negligibly small compared to the diagonal one \cite{bauer}. This
is also true for the $\omega$ meson photoproduction at high energy
\cite{bauer, sibir1}.
As
illustrated by Sibirstev et al. \cite{sibir1}, the low energy $\omega$ meson
photoproduction data is better understood by the off-diagonal process, i.e.,
$\gamma \to \rho^0$; $\rho^0 p \to \omega p$.
More
precisely, the reaction $\rho^0 p \to \omega p$ proceeds due to one pion
exchange (OPE) interaction \cite{bauer, frim}, i.e., the $\omega$ meson
photoproduction in the off-diagonal process arises because of the pion
exchange interaction.
The
superscripts $``D"$ and $``O"$ are used afterwards to differentiate the
vector meson production due to diagonal and off-diagonal processes of VMD
model respectively.

The differential cross section $\frac{d\sigma}{dm}$ of the $e^+e^-$
invariant mass $m$ distribution in the $\gamma p \to pe^+e^-$ reaction is
calculated in the multi-GeV region using both diagonal and off-diagonal
processes of VMD model.
The
interference of these processes cannot exist \cite{bauer}, since the
previous process involves natural parity exchange whereas the latter
(because of OPE interaction) occurs due to unnatural parity exchange.
Therefore, the total
differential cross section $\frac{d\sigma}{dm}$ of the above reaction
can be written as
\begin{equation}
\frac{d\sigma}{dm} = \frac{d\sigma^D}{dm} 
                   + \frac{d\sigma^O}{dm},
\label{dcrs}
\end{equation}
where $\frac{d\sigma^D}{dm}$ represents the differential cross section of
the $\gamma p \to Vp; ~V \to e^+e^-$ reaction due to the diagonal process,
and $\frac{d\sigma^O}{dm}$ denotes that of the $\gamma p \to \omega p$; ~
$\omega \to e^+e^-$ reaction due to off-diagonal process of VMD model.

\section{Formalism}

The production of $V$(vector) meson in the $\gamma p \to V p$ reaction is
described by the production amplitude $\Pi_{\gamma p \to Vp} ({\bf r})$,
i.e.,
\begin{equation}
\Pi_{\gamma p \to Vp} ({\bf r}) = K f_{\gamma p \to Vp} (q^2=0)
\delta ({\bf r});
\label{pia}
\end{equation}
$ K = -4\pi (1 + E_V/E_{p^\prime}) $ \cite{das2, pautz}. $E_{p^\prime}$ is
the energy of proton in the final state. $f_{\gamma p \to Vp} (q^2=0)$
denotes the forward amplitude of the $\gamma p \to Vp$ reaction in the
$\gamma p$ cm system. $q^2$ is the four-momentum transfer from the
projectile. $\delta ({\bf r})$ is the density distribution of proton
(point particle).

The vector meson produced in the $\gamma p$ reaction propagates
in the free space before it decays into $e^+e^-$. The propagator of this
meson is given by $ (-g^\mu_\nu + \frac{1}{m^2} k^\mu_V k_{V,\nu} )
G_{0V} (m) $ \cite{das2}. The scalar part of it, i.e., $G_{0V}(m)$, can be
written as
\begin{equation}
G_{0V}(m) = \frac{1}{m^2-m^2_V+im_V\Gamma_V(m)}, 
\label{vmp}
\end{equation}
where
$m$ is the vector meson mass, i.e., $e^+e^-$ invariant mass. $m_V$ is the
pole mass of the meson: $m_{\rho^0}=775.26$ MeV and $m_{\omega}=782.65$ MeV,
as listed in Ref.~\cite{olive}.
The
total decay width of the vector meson $\Gamma_V(m)$ is composed of the
partial widths of various decay channels
\cite{olive}:
$ \Gamma_{\rho^0} (m) =
99.93\% \Gamma_{\rho^0 \to \pi^+\pi^-} (m) +
0.06\%  \Gamma_{\rho^0 \to \pi^0\gamma} (m)  +
0.01\%  \Gamma_{\rho^0 \to e^+e^-} (m) $,
and
$ \Gamma_\omega (m) =
89.9\% \Gamma_{\omega \to \pi^+\pi^-\pi^0} (m) +
8.28\% \Gamma_{\omega \to \pi^0\gamma} (m) +
1.53\% \Gamma_{\omega \to \pi^+\pi^-} (m)  +
0.29\% \Gamma_{\omega \to e^+e^-} (m) $.
The partial decay widths are illustrated in Ref. \cite{das3} except the
dielectron decay width, i.e., $ \Gamma_{V \to e^+e^-} (m) $, which is given
by
\cite{sibir1}
\begin{equation}
\Gamma_{V \to e^+e^-} (m)  \simeq  \frac{ \pi }{ 3 }
\left ( \frac{ \alpha_{em} }{ \gamma_{\gamma V} } \right )^2 m.
\label{dew}
\end{equation}

The differential cross section $\frac{d\sigma}{dm}$ for the dilepton
invariant mass $m$ distribution in the ($\gamma, V \to e^+e^-$) reaction
on proton can be written as
\begin{equation}
\frac{d\sigma}{dm} = \int d\Omega_V P_{e^+e^-}
\left | \sum_{V=\rho^0,\omega}
        \Gamma^{1/2}_{V \to e^+e^-} (m) F_{\gamma, V} \right |^2;
\label{dx1}
\end{equation}
$F_{\gamma, V}= G_V (m) K f_{\gamma p \to Vp} (q^2=0)$.
$\Omega_V$
is the solid angle subtended by the vector meson momentum ${\bf k}_V$.
The
factor $P_{e^+e^-}$ arises because of the phase space of the reaction:
$ P_{e^+e^-} = \frac{3}{8(2\pi)^3}
\frac{k^2_Vm^2}{m_pk_\gamma |k_V E_i - k_\gamma cos\theta_V E_V|} $.
$P_{e^+e^-}$ and $\Omega_V$ are same for both $\rho^0$ and $\omega$ mesons.

$\frac{d\sigma}{dm}$ in Eq.~(\ref{dx1}) has been calculated using the vector
meson photoproduction amplitude $f_{\gamma N \to V N}$.
Both
diagonal and off-diagonal processes of the vector meson dominance (VMD)
model relates $f_{\gamma N \to V N}$ to the vector meson nucleon scattering
amplitude $f_{V^\prime N \to V N}$ \cite{sibir1} as
\begin{equation}
 f_{\gamma N \to V N}
= \sum_{V^\prime = \rho^0, \omega, ...}
  \frac{\sqrt{\pi \alpha_{em}}}{\gamma_{\gamma V^\prime}}
   f_{V^\prime N \to V N},
\label{famp}
\end{equation}
where the constants $\alpha_{em}$ and $\gamma_{\gamma V^\prime}$ are
defined in Eq.~(\ref{wgma}).
$\gamma_{\gamma V^\prime}$
for the $\omega$ meson, i.e., $\gamma_{\gamma \omega}$, is $\sim 3.5$ times
larger than $\gamma_{\gamma \rho}$. Therefore, the photoproduction amplitude
of the $\rho^0$ meson can be described well by the diagonal process
\cite{bauer}, i.e.,
$ f_{\gamma N \to \rho^0 N}  \approx  f^D_{\gamma N \to \rho^0 N}
= \frac{\sqrt{\pi \alpha_{em}}}{\gamma_{\gamma \rho}}
f_{\rho^0 N \to \rho^0 N} $.
The
$\rho^0$ meson nucleon scattering amplitude $f_{\rho^0 N \to \rho^0 N}$
(extracted from the elementary $\rho^0$ meson photoproduction data)
is taken from the calculation of Kondratyuk et al., \cite{kon}.

The $\omega$ meson production in the $\gamma p \to \omega p$ reaction, as
mentioned earlier, is illustrated by both diagonal and off-diagonal
processes of VMD model. Therefore, the amplitude of this reaction,
according to Eq.~(\ref{famp}), is
$f_{\gamma N \to \omega N} = f^D_{\gamma N \to \omega N}
 + f^O_{\gamma N \to \omega N}$,
where
$f^D_{\gamma N \to \omega N} =
\frac{\sqrt{\pi \alpha_{em}}}{\gamma_{\gamma \omega}}
f_{\omega N \to \omega N}$  (diagonal process)
and
$f^O_{\gamma N \to \omega N} =
\frac{\sqrt{\pi \alpha_{em}}}{\gamma_{\gamma \rho}}
f_{\rho^0 N \to \omega N}$  (off-diagonal process).
The
energy dependent differential cross section of the previous process
\cite{bauer} is given by
\begin{equation}
\frac{d\sigma^D}{dq^2} (\gamma p \to \omega p) |_{q^2=0}
= C \left ( 1 + \frac{D}{E_\gamma} \right );
\label{ftD1}
\end{equation}
$C=9.3$ $\mu$b/GeV$^2$ and $D=1.4$ GeV. According to VMD model
\cite{sibir1, lyka}, it
can be written as
\begin{equation}
\frac{d\sigma^D}{dq^2} (\gamma p \to \omega p) |_{q^2=0}
= \frac{\alpha_{em}}{16\gamma^2_{\gamma \omega}}
  \left ( \frac{ \tilde{k}_\omega }{ \tilde{k}_\gamma } \right )^2
  \ [1+\alpha^2_{\omega N}] (\sigma^{\omega N}_t)^2,
\label{ftD2}
\end{equation}
where $\tilde{k}_\omega$ and $\tilde{k}_\gamma$ are the c.m. momenta in
the $\omega N$ and $\gamma N$ systems respectively, evaluated at the c.m.
energy of $\gamma N$ system.
$\alpha_{\omega N}$
is the ratio of the real to imaginary part of the $\omega N$ scattering
amplitude, taken from Ref.~\cite{sibir3}. $\sigma^{\omega N}_t$ is the
total $\omega N$ scattering cross section.
Using
Eqs.~(\ref{famp})-(\ref{ftD2}), the energy dependent values of the $\omega$
meson photoproduction amplitude $f^D_{\gamma p \to \omega p}$ has been
evaluated.

Friman et al., \cite{frim} have calculated the total cross section
$\sigma^O_t$ and the differential cross section for the four-momentum
transfer distribution $\frac{d\sigma^O}{dq^2}$ of the
$\gamma p \to \omega p$ reaction using the off-diagonal process of VMD
model,
i.e.,
$\gamma \to \rho^0; ~\rho^0 p \to \omega p$. They have used one pion
exchange interaction to describe the $\rho^0 p \to \omega p$ reaction.
The calculated results due to them reproduced well the data of both
$\sigma^O_t$ and $\frac{d\sigma^O}{dq^2}$ at low energy.
Therefore,
$\frac{d\sigma^O}{dq^2} (\gamma p \to \omega p)$ calculated by Friman
et al. \cite{frim} is used to extract $f^O_{\gamma p \to \omega p} (q^2=0)$.
As
illustrated by Sibirtsev et al. \cite{sibir1}, the forward photoproduction
cross section $\frac{d\sigma}{dq^2}|_{q^2=0}$ of the $\gamma p \to \omega p$
reaction is determined through the extrapolation of $\frac{d\sigma}{dq^2}$
to $q^2=0$.
This
extrapolation is done over a range of $q^2$ (close to $q^2=0$, e.g.,
$q^2=$ -0.01 to -0.03 GeV$^2$) where a fit of the form
$\frac{d\sigma}{dq^2} = \frac{d\sigma}{dq^2}|_{q^2=0} exp(bq^2)$ can be used.
Due
to this reason, $\frac{d\sigma}{dq^2}|_{q^2=0}$ does not vanish at low
energy where the differential cross section is clearly dominated by one
pion exchange interaction.
The
magnitude of $f^O_{\gamma p \to \omega p} (q^2=0)$ is evaluated using the
relation \cite{sibir1}:
\begin{equation}
\frac{d\sigma^O}{dq^2} (\gamma p \to \omega p)|_{q^2=0}
= \frac{\pi}{k^2_\gamma} | f^O_{\gamma p \to \omega p} (q^2=0) |^2.
\label{xfmd}
\end{equation}
In
fact, $| f^O_{\gamma p \to \omega p} (q^2=0)|^2$ is required to calculate
the cross section $\frac{d\sigma^O}{dm}$ in Eq.~(\ref{dx1}).

\section{Result and Discussions}

\begin{figure}
\begin{center}
\centerline {\vbox {
\psfig{figure=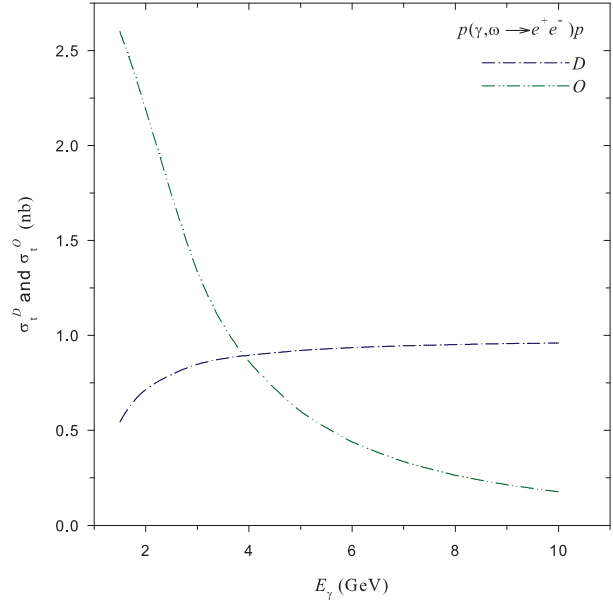,height=08.0 cm,width=08.0 cm}
}}
\caption{
(color online).
The beam energy dependent total cross sections $\sigma_t^D$ and
$\sigma_t^O$ of the $\gamma p \to \omega p; ~\omega \to e^+e^-$ reaction.
The superscripts ``$D$'' and ``$O$'' represent the diagonal and off-diagonal
processes of the vector meson dominance (VMD) model respectively (see the
detail in the text).
}
\label{FOm}
\end{center}
\end{figure}

Using Eq.~(\ref{dx1}), the total cross section of the
$\gamma p \to \omega p$; ~ $\omega \to e^+e^-$ reaction, i.e.,
$\sigma_t = \int (\frac{d\sigma}{dm}) dm$, is calculated.
For
the diagonal process, $\sigma_t^D$ is calculated using
$f^D_{\gamma p \to \omega p}$ described in Eq.~(\ref{ftD2})
whereas
that due to off-diagonal process (i.e., $\sigma_t^O$) is worked out using
$f^O_{\gamma p \to \omega p}$ in Eq.~(\ref{xfmd}).
The
calculated results are compared in Fig.~\ref{FOm} for the beam energy
$E_\gamma$ range $1.5-10$ GeV. $\sigma_t^D$ is shown by the dot-dashed
curve which increases significantly with beam energy upto $E_\gamma =4$ GeV.
Beyond that, $\sigma_t^D$ increases (with $E_\gamma$) very slowly.
$\sigma_t^O$ (presented by the dot-dot-dashed curve) falls with $E_\gamma$.
This
figure distinctly shows that $\sigma_t^O$ is significantly larger than
$\sigma_t^D$ at low energy, i.e., $E_\gamma <4$ GeV.
For
the exclusive $\omega$ meson production in the intermediate state of the
$\gamma p \to pe^+e^-$ reaction, $\sigma_t^D$ increases and $\sigma_t^O$
decreases with the increase in the beam energy.
Fig.~\ref{FOm}
shows that both diagonal and off-diagonal processes are significant for
the $\omega$ meson photoproduction reaction around the beam energy
$E_\gamma = 4$ GeV. $\sigma_t^D$ is distinctly dominant over
$\sigma_t^O$ at $E_\gamma \simeq 10$ GeV.

As mentioned earlier, the diagonal process of VMD model describes the
production of both $\rho^0$ and $\omega$ mesons in the $\gamma p$ reaction.
Therefore, the photoproduction amplitudes of these mesons, i.e.,
$f^D_{\gamma p \to \rho^0 p}$
and
$f^D_{\gamma p \to \omega p}$, due to the above process are added coherently
to evaluate the differential cross section for the dielectron mass $m$
distribution (i.e., $\frac{d\sigma^D}{dm}$) of the $\gamma p \to Vp$;
~$V \to e^+e^-$ reaction.
The
calculated results for $\frac{d\sigma^D}{dm}$ at $E_\gamma =1.5$ GeV are
shown in Fig.~\ref{Fint1p5}(a). The short-dashed curve represents
$\frac{d\sigma^D}{dm}$ (calculated using $f^D_{\gamma p \to \rho^0 p}$)
occurring due to $\rho^0$ meson.
The
broad distribution of this curve arises because of the large width
($\sim 150$ MeV) of the $\rho^0$ meson.
The
narrow distribution shown by the dot-dashed curve (peaking at the $\omega$
meson pole mass, i.e., $\sim 782$ MeV) denotes $\frac{d\sigma^D}{dm}$
(calculated using $f^D_{\gamma p \to \omega p}$) because of the $\omega$
meson.
The
dotted curve (showing sharp peak at $\sim 782$ MeV) originates due to the
$\rho -\omega$ interference.
Unless
mentioned explicitly, $\frac{d\sigma^D}{dm}$ represents the cross section
due to the coherent contribution of $\rho^0$ and $\omega$ mesons. It is
illustrated by the large-dashed curve,
which
shows narrow peak at the $\omega$ meson mass $\sim 782$ and broad base
because of the large width of the $\rho$ meson.
As
shown in this figure, the cross section is increased considerably (i.e., by
a factor $\approx 2$) due to the $\rho -\omega$ interference.

In Fig.~\ref{Fint1p5}(b), the differential cross sections
$\frac{d\sigma^D}{dm}  (\gamma p \to V p; ~V \to e^+e^-)$
(large-dashed curve) and
$\frac{d\sigma^O}{dm}  (\gamma p \to \omega p; ~\omega \to e^+e^-)$
(dot-dot-dashed curve) are compared at $E_\gamma =1.5$ GeV.
The
previous is illustrated in Fig.~\ref{Fint1p5}(a). Since the latter describes
only the $\omega$ meson production, it shows narrow distribution peaking
at $\sim 782$ MeV.
This
figure illustrates that $\frac{d\sigma^O}{dm}$ is significantly larger than
$\frac{d\sigma^D}{dm}$, though the latter (as described in
Fig.~\ref{Fint1p5}(a)) is remarkably increased because of the
$\rho -\omega$ interference.
The
$e^+e^-$ emission in the $\gamma p$ reaction dominantly occurs due to the
decay of $\omega$ meson produced (in the intermediate state) because of the
off-diagonal process of VMD model.
The
total differential cross section $\frac{d\sigma}{dm}$, i.e.,
$\frac{d\sigma^D}{dm} + \frac{d\sigma^O}{dm}$ in Eq.~(\ref{dcrs}), is
described by the solid curve in Fig.~\ref{Fint1p5}(b). It shows the feature
similar to other curves appearing in the figure.

The differential cross sections of the $\gamma p \to pe^+e^-$ reaction
calculated at $E_\gamma =3$ GeV are shown in Fig.~\ref{Fint3}. Various
curves appearing in this figure (and all other figures presented afterwards)
illustrate those as stated in Fig.~\ref{Fint1p5}.
The
calculated results shown in Fig.~\ref{Fint3}(a) is qualitatively similar to
those in Fig.~\ref{Fint1p5}(a).
Unlike
to that in Fig.~\ref{Fint1p5}(b), Fig.~\ref{Fint3}(b) distinctly shows that
$\frac{d\sigma^D}{dm}$ supersedes $\frac{d\sigma^O}{dm}$ because of the
$\rho -\omega$ interference.
$\frac{d\sigma^D}{dm}$ increases and $\frac{d\sigma^O}{dm}$
decreases with the increase in the beam energy.
The
latter (i.e., $\frac{d\sigma^O}{dm}$) is significantly large at 3 GeV, as
the total differential cross section $\frac{d\sigma}{dm}$ in
Eq.~(\ref{dcrs}) is increased by $\sim 60\%$ due to it.
The
calculated results (not presented) show that the enhancement in
$\frac{d\sigma}{dm}$ because of $\frac{d\sigma^O}{dm}$ is $\sim 25\%$ at
5 GeV.

To look for the above features at higher energy, the differential cross
sections $\frac{d\sigma^D}{dm}$ and $\frac{d\sigma^O}{dm}$ are calculated
at $E_\gamma$ taken equal to 10 GeV, and those are presented in
Fig.~\ref{Fint10}. The enhancement in the cross section due to
$\rho -\omega$ interference is distinctly visible in Fig.~\ref{Fint10}(a).
Unlike
to those in Figs.~\ref{Fint1p5}(b) and \ref{Fint3}(b), Fig.~\ref{Fint10}(b)
shows that $\frac{d\sigma^O}{dm}$ is negligibly small compared to
$\frac{d\sigma^D}{dm}$, i.e., the differential cross section
$\frac{d\sigma}{dm}$ in Eq.~(\ref{dcrs}) is increased only $6.7\%$ due to
$\frac{d\sigma^O}{dm}$ at this energy.
Therefore,
the dilepton $e^+e^-$ emission in the $\gamma p$ reaction at $E_\gamma =10$
GeV can be described well by the diagonal process of VMD model.

The calculated results illustrate that the $\gamma p \to pe^+e^-$ reaction
at the fixed photon beam energy in the region $E_\gamma > 1.5$ GeV to
$E_\gamma \sim 5$ GeV occurs due to both diagonal and off-diagonal processes
of VMD model.
Since
the latter process (as mentioned earlier) arises due to one pion exchange
interaction, it is significant in the quoted energy region.
At
$E_\gamma \sim 10$ GeV, the considered reaction can be accounted well by
the diagonal process which does not involve the pion exchange interaction.
Therefore, this interaction around 10 GeV is insignificant for the
considered reaction.

The dilepton invariant mass distribution in the $\gamma p \to pe^+e^-$
reaction was measured at JLab using Bremsstrahlung photon beam
which
possesses a certain range of energy. The Bremsstrahlung cross section
$\sigma_B$ varies as $\frac{1}{E_\gamma}$ \cite{sober}. Therefore, the
differential cross section $\frac{d\sigma}{dm}$ of the $e^+e^-$ invariant
mass $m$ distribution in the Bremsstrahlung beam induced
$\gamma p \to pe^+e^-$ reaction \cite{kask} can be written as
\begin{equation}
\frac{d\sigma}{dm} = 
\int^{E_{\gamma, mx}}_{E_{\gamma, mn}} dE_\gamma
W(E_\gamma) \frac{d\sigma}{dm} (E_\gamma), 
\label{bxsn}
\end{equation}
with $W(E_\gamma) \propto \frac{1}{E_\gamma}$. In this equation,
$\frac{d\sigma}{dm} (E_\gamma)$ denotes the differential cross section at
the fixed beam energy, given in Eq.~(\ref{dx1}). $E_{\gamma, mn}$ and
$E_{\gamma, mx}$ are the minimum and maximum energies respectively of the
Bremsstrahlung photon beam.

The differential cross section of the $\gamma p \to pe^+e^-$ reaction has
been calculated using Eq.~(\ref{bxsn}) for the Bremsstrahlung photon beam
energy range $E_{\gamma, mn} =1.2$ GeV to $E_{\gamma, mx} =5.4$ GeV.
This
is the energy range used for the measurement done at Jlab, and the
preliminary results (as mentioned earlier) are reported in Ref.~\cite{djal}.
The
calculated results are presented in Fig.~\ref{FBb1}, and various curves
appearing in it are described in Fig.~\ref{Fint1p5}. Fig.~\ref{FBb1}(a)
shows $\frac{d\sigma^D}{dm}$ in the considered energy range is significantly
increased due to $\rho -\omega$ interference.
Both
cross sections $\frac{d\sigma^D}{dm}$ and $\frac{d\sigma^O}{dm}$, as
illustrated in Fig.~\ref{FBb1}(b), are significantly large for the
quoted reaction, i.e.,
the cross section $\frac{d\sigma}{dm}$
(= $\frac{d\sigma^D}{dm} + \frac{d\sigma^O}{dm}$ in Eq.~(\ref{dcrs}))
is increased by $\sim 72\%$ because of $\frac{d\sigma^O}{dm}$.

The upgrade accelarator facility at JLab can provide electron beam of
energy 12 GeV. Therefore, the upper limit of the energy of Bremsstrahlung
photon beam can be considered $\sim 10$ GeV. Bremsstrahlung cross section
$\sigma_B$ at this energy is much less than that at 5.4 GeV because it falls
as $\frac{1}{E_\gamma}$ (mentioned above).
As
shown in Fig.~\ref{Fint10}(b), the differential cross section
$\frac{d\sigma^O}{dm}$ of the $\gamma p \to pe^+e^-$ reaction is negligibly
small (compared to $\frac{d\sigma^D}{dm}$) at $E_\gamma =10$ GeV.
Both
$\sigma_B$ ($ \propto \frac{1}{E_\gamma}$) and $\frac{d\sigma^O}{dm}$
(see Fig.~\ref{Fint1p5}(b)) at $E_\gamma << 10$ GeV are much larger than
those at 10 GeV.
Since
the energy of Bremsstrahlung photon beam possesses a certain range, the
differential cross sections are calculated to search the contribution of
$\frac{d\sigma^O}{dm}$ to $\frac{d\sigma}{dm}$ for the beam energy range
$E_{\gamma, mn} =1.2$ GeV to $E_{\gamma, mx} =10$ GeV.
The
calculated results presented in Fig.~\ref{FBb2} show that they are
qualitatively similar to those depicted in Fig.~\ref{FBb1}.
The
figure \ref{FBb2}(b) shows that $\frac{d\sigma^O}{dm}$ in the quoted energy
range is considerably large as $\frac{d\sigma}{dm}$ is increased by
$\sim 43\%$ due to it.
Therefore,
the pion exchange interaction is significant for the reactions illustrated
in Figs.~\ref{FBb1} and \ref{FBb2}.

\section{Conclusions}

The dielectron emission in the  $\gamma p$ reaction has been studied in
the multi-GeV region using the vector meson dominance (VMD) model. The
$e^+e^-$ in the final state is assumed to arise because of the decay of
the vector meson $V$ (i.e., either $\rho^0$ and $\omega$ mesons) produced
in the intermediate state, i.e., $\gamma p \to Vp; ~V \to e^+e^-$.
The
photoproduction of both $\rho^0$ and $\omega$ mesons has been described
by the diagonal process of VMD model (i.e., $\gamma \to V$; ~$Vp \to Vp$).
Therefore, the $\rho -\omega$ interference occurs in this process.
The
$\omega$ meson is also photoproduced due to off-diagonal process of VMD
model,
i.e., $\gamma \to \rho^0; ~\rho^0 p \to \omega p$. The reaction
$\rho^0 p \to \omega p$ proceeds because of one pion exchange interaction,
i.e., the off-diagonal process arises due to this interaction. The
interference between the diagonal and off-diagonal processes cannot exist.

The calculated results show that the cross section of the
$\gamma p \to pe^+e^-$ reaction due to the diagonal process
($\frac{d\sigma^D}{dm}$) increases whereas that due to off-diagonal process
($\frac{d\sigma^O}{dm}$) decreases with the increase in the beam energy.
In
the region
$E_\gamma >1.5$ GeV to $E_\gamma \sim 5$ GeV, the considered reaction
occurs because of both diagonal and off-diagonal processes. Therefore, the
pion exchange interaction is significant in this region.
At
higher energy (i.e., $E_\gamma \sim 10$ GeV), $\frac{d\sigma^O}{dm}$ is
negligibly small compared to $\frac{d\sigma^D}{dm}$, i.e., the pion
exchange interaction is insignificant around this energy.

For Bremsstrahlung photon beam of minimum energy $E_{\gamma, mn} =1.2$ GeV,
$\frac{d\sigma^O}{dm}$ of the $\gamma p \to pe^+e^-$ reaction is found
considerably large irrespective of the maximum beam energy $E_{\gamma, mx}$
taken equal to 5.4 or 10 GeV.
Therefore,
the pion exchange interaction is significant for the Bremsstrahlung beam
induced $\gamma p \to pe^+e^-$ reaction in the above energy regions.

\section{Acknowledgement}
The author thanks the anonymous referee for the comments which helped
to improve the quality of the paper.

\newpage

\begin{figure}[h]
\begin{center}
\centerline {\vbox {
\psfig{figure=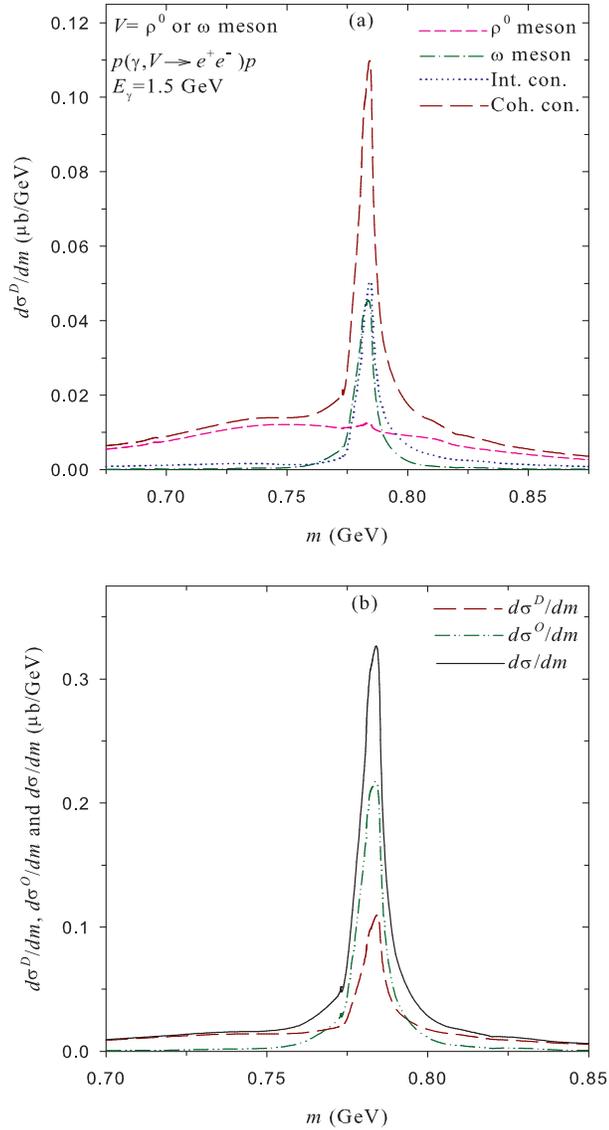,height=15.0 cm,width=08.0 cm}
}}
\caption{
(color online).
(a) $\frac{d\sigma^D}{dm}$ calculated at
$E_\gamma =1.5$ GeV using the diagonal process of VMD model.
$\frac{d\sigma^D}{dm}$ due to $\rho$ and $\omega$ mesons are represented
by the short-dashed and dot-dashed curves respectively.
The
dotted curve denotes that because of the $\rho^0 -\omega$
interference. $\frac{d\sigma^D}{dm}$ due to the coherence contribution of
these mesons is shown by the large-dashed curve.
(b)
$\frac{d\sigma^D}{dm}$ and $\frac{d\sigma^O}{dm}$ are compared. The latter
(shown by the dot-dot-dashed curve)
is calculated for the $\omega$ meson produced in the off-diagonal process of
VMD model.
$ \frac{d\sigma}{dm} = \frac{d\sigma^D}{dm} + \frac{d\sigma^O}{dm} $ is
represented by the solid curve in the figure.
}
\label{Fint1p5}
\end{center}
\end{figure}

\begin{figure}[h]
\begin{center}
\centerline {\vbox {
\psfig{figure=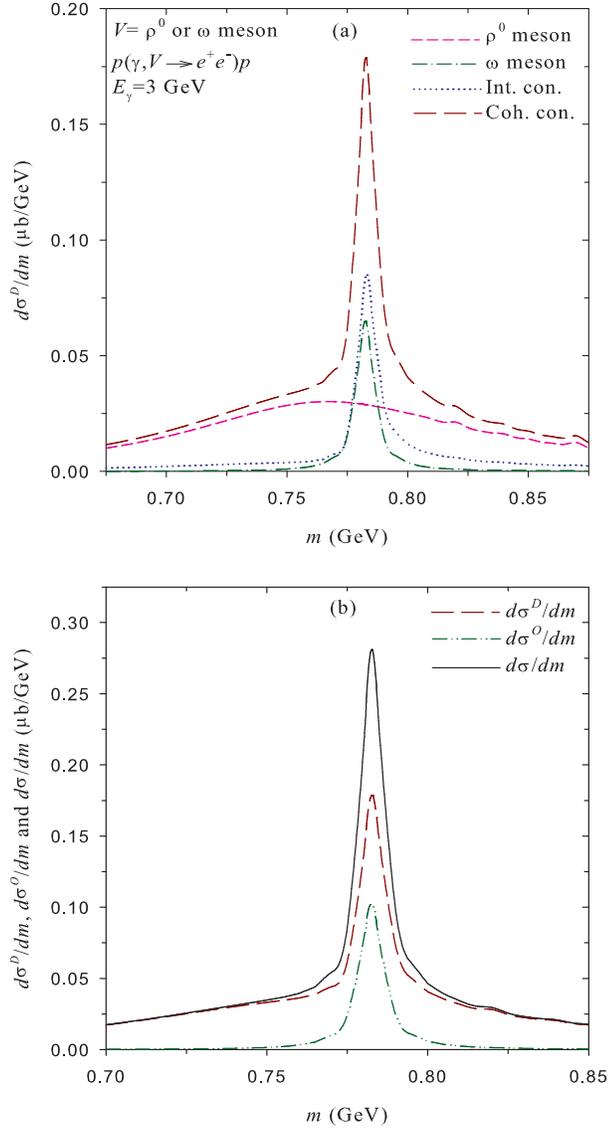,height=15.0 cm,width=08.0 cm}
}}
\caption{
(color online).
Same as those presented in Fig.~\ref{Fint1p5} but for the beam energy
$E_\gamma$ taken equal to 3 GeV.
}
\label{Fint3}
\end{center}
\end{figure}

\begin{figure}[h]
\begin{center}
\centerline {\vbox {
\psfig{figure=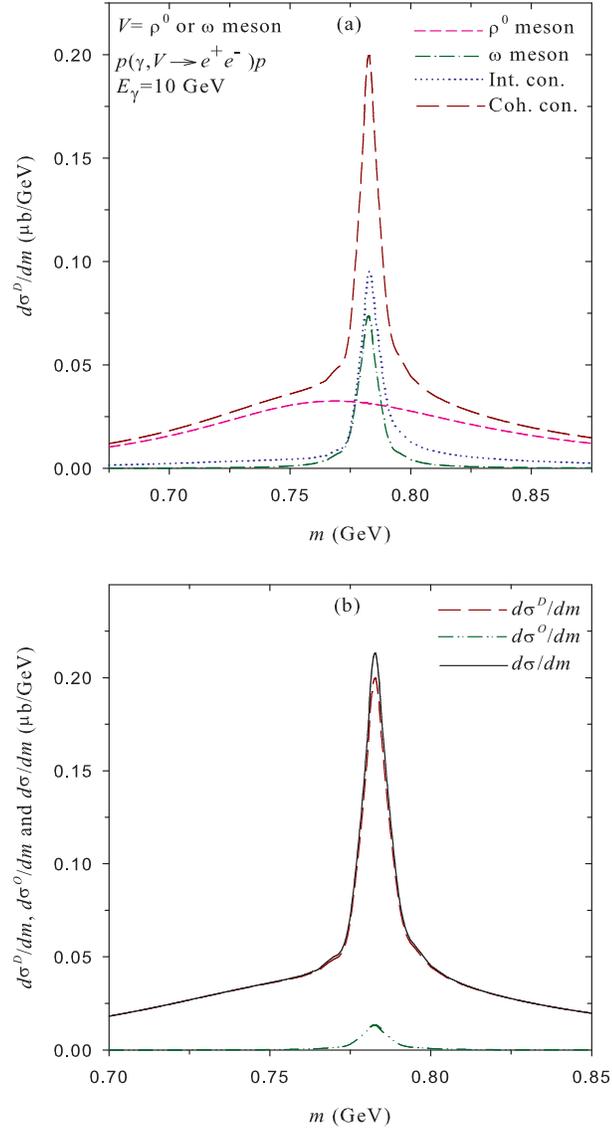,height=15.0 cm,width=08.0 cm}
}}
\caption{
(color online).
Same as those illustrated in Fig.~\ref{Fint1p5} but for $E_\gamma =10$ GeV.
}
\label{Fint10}
\end{center}
\end{figure}

\begin{figure}[h]
\begin{center}
\centerline {\vbox {
\psfig{figure=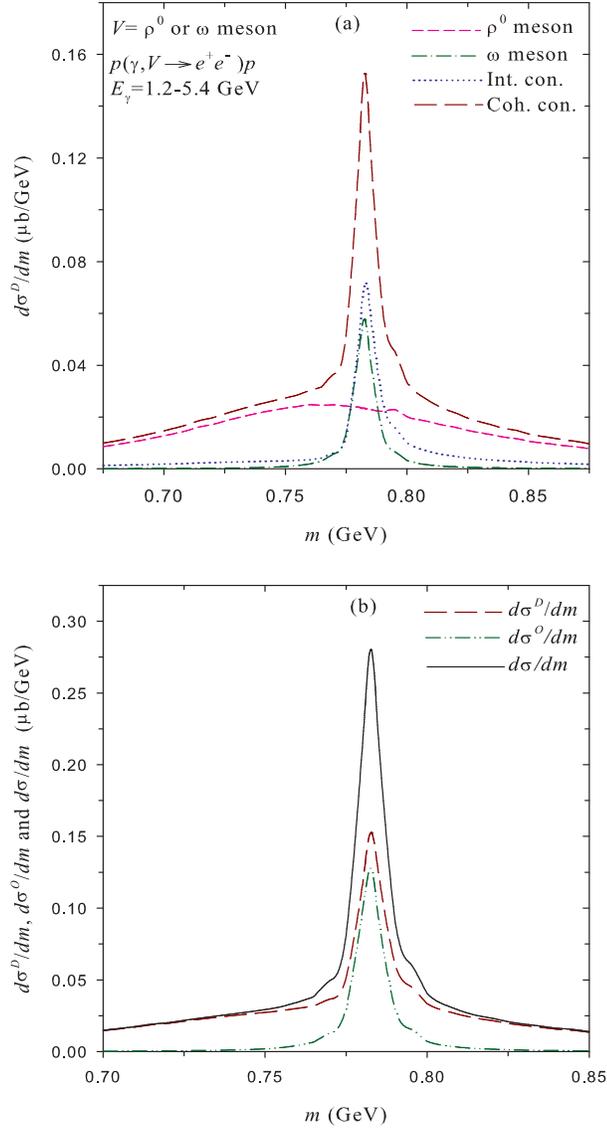,height=15.0 cm,width=08.0 cm}
}}
\caption{
(color online).
The curves represent those as described in Fig.~\ref{Fint1p5} but for the
Bremsstrahlung photon beam of the energy range: $E_\gamma = 1.2 - 5.4$ GeV.
}
\label{FBb1}
\end{center}
\end{figure}

\begin{figure}[h]
\begin{center}
\centerline {\vbox {
\psfig{figure=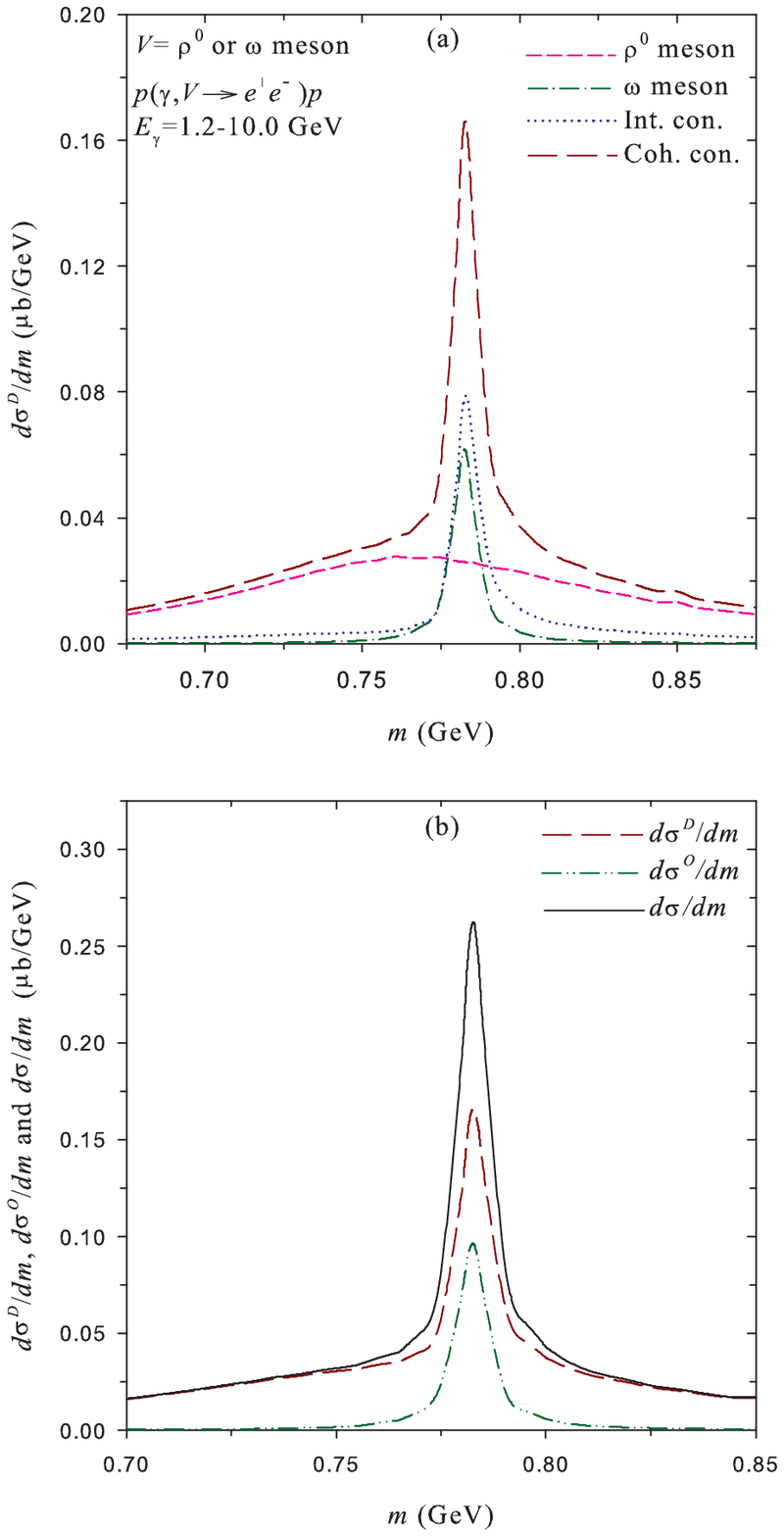,height=15.0 cm,width=08.0 cm}
}}
\caption{
(color online).
Same as those in Fig.~\ref{FBb1} but for $E_\gamma =1.2-10$ GeV.
}
\label{FBb2}
\end{center}
\end{figure}

{\bf Figure Captions}
\begin{enumerate}

\item
(color online).
The beam energy dependent total cross sections $\sigma_t^D$ and
$\sigma_t^O$ of the $\gamma p \to \omega p; ~\omega \to e^+e^-$ reaction.
The superscripts ``$D$'' and ``$O$'' represent the diagonal and off-diagonal
processes of the vector meson dominance (VMD) model respectively (see the
detail in the text).

\item
(color online).
(a) $\frac{d\sigma^D}{dm}$ calculated at
$E_\gamma =1.5$ GeV using the diagonal process of VMD model.
$\frac{d\sigma^D}{dm}$ due to $\rho$ and $\omega$ mesons are represented
by the short-dashed and dot-dashed curves respectively.
The
dotted curve denotes that because of the $\rho^0 -\omega$
interference. $\frac{d\sigma^D}{dm}$ due to the coherence contribution of
these mesons is shown by the large-dashed curve.
(b)
$\frac{d\sigma^D}{dm}$ and $\frac{d\sigma^O}{dm}$ are compared. The latter
(shown by the dot-dot-dashed curve)
is calculated for the $\omega$ meson produced in the off-diagonal process of
VMD model.
$ \frac{d\sigma}{dm} = \frac{d\sigma^D}{dm} + \frac{d\sigma^O}{dm} $ is
represented by the solid curve in the figure.

\item
(color online).
Same as those presented in Fig.~\ref{Fint1p5} but for the beam energy
$E_\gamma$ taken equal to 3 GeV.

\item
(color online).
Same as those illustrated in Fig.~\ref{Fint1p5} but for $E_\gamma =10$ GeV.

\item
(color online).
The curves represent those as described in Fig.~\ref{Fint1p5} but for the
Bremsstrahlung photon beam of the energy range: $E_\gamma = 1.2 - 5.4$ GeV.

\item
(color online).
Same as those in Fig.~\ref{FBb1} but for $E_\gamma =1.2-10$ GeV.

\end{enumerate}

\newpage

\newpage

\newpage

\newpage

\newpage

\newpage


\newpage
\begin{thebibliography}{100}

\bibitem{bauer}
T. H. Bauer, R. D. Spital, D. R. Yennie and F. M. Pipkin,
Rev. Mod. Phys. {\bf 50} (1978) 261;
Erratum, ibid, {\bf 51} (1979) 407.

\bibitem{behr0}
H.-J. Behrend et al., Phys. Rev. Lett. {\bf 24} (1970) 336;
H. Alvensleben et al., Nucl. Phys. B {\bf 18} (1970) 333.

\bibitem{conn}
H. B. O'Connell, B. C. Pearce, A. W. Thomas and A. G. Williams,
Prog. Part. Nucl. Phys. {\bf 39} (1997) 201.

\bibitem{biggs2}
P. J. Biggs et al., Phys. Rev. Lett. {\bf 24} (1970) 1201;
H.-J. Behrend et al., Phys. Rev. Lett. {\bf 27} (1971) 61;
H. Alvensleben et al., Phys. Rev. Lett. {\bf 27} (1971) 888.

\bibitem{das1}
Swapan Das, Phys. Rev. C {\bf 94} (2016) 025204.

\bibitem{alven}
H. Alvensleben et al., Phys. Rev. Lett. {\bf 25} (1970) 1373.

\bibitem{biggs}
P. J. Biggs et al., Phys. Rev. Lett. {\bf 24} (1970) 1197.

\bibitem{lutz}
M. F. M. Lutz and M. Soyeur, Nucl. Phys. A {\bf 760} (2005) 85.

\bibitem{djal}
C. Djalali et al., Proceeding of Science (Hadron 2013) 176.

\bibitem{saku}
J. J. Sakurai, Currents and Mesons (The University of Chicago Press,
Chicago, 1969);
R. K. Bhaduri, Models of the Nucleon (Addison-Wesley Publishing Company,
Inc., California, 1988).

\bibitem{olive}
K. A. Olive et al., (Particle Data Group),
Chin. Phys. C {\bf 38} (2014) 090001.

\bibitem{sibir1}
A. Sibirtsev, H.-W. Hammer, U.-G. Mei$\ss$ner and A. W. Thomas,
Eur. Phys. J. A {\bf 29} (2006) 209;
A. Sibirtsev, H.-W. Hammer and U.-G. Mei$\ss$ner,
Eur. Phys. J. A {\bf 37} (2008) 287.

\bibitem{frim}
B. Friman and M. Soyeur
Nucl. Phys. A {\bf 600} (1996) 477.

\bibitem{das2}
S. Das, Eur. Phys. J. A {\bf 49} (2013) 123; Pramana {\bf 75} (2010) 665.

\bibitem{pautz}
A. Pautz and G. Shaw, Phys. Rev. C {\bf 57} (1998) 2648.

\bibitem{das3}
S. Das, Rev. C {\bf 72} (2005) 064619;
W. Peters, et al., Nucl. Phys. A {\bf 632} (1998) 109;
S. Das, Rev. C {\bf 78} (2008) 045210.

\bibitem{kon}
L. A. Kondratyuk, A. Sibirtsev, W. Cassing, Ye. S. Golubeva
and M. Effenberger, Phys. Rev. C {\bf 58} (1998) 1078.

\bibitem{lyka}

G. I. Lykasov, W. Cassing, A. Sibirtsev and M. V. Rzjanin,
Eur. Phys. J. A {\bf 6} (1999) 71.

\bibitem{sibir3}
A. Sibirtsev, Ch. Elster and J. Speth, arXiv:nucl-th/0203044.

\bibitem{sober}
D. I. Sober, et al., Nucl. Instr. and Meth. A {\bf 440} (2000) 263.

\bibitem{kask}
M. Kaskulov, E. Hernandez and E. Oset, Eur. Phys. J. A {\bf 31} (2007) 245;
S. Das, Phys. Rev. C {\bf 83} (2011) 064608.


\end{thebibliography}
\end{document}